\documentclass[twocolumn]{aastex631}

\begin{document}

\title{
Statistics for Galaxy Outflows at $z\sim6-9$ with Imaging and Spectroscopic Signatures\\ 
Identified with JWST/NIRCam and NIRSpec Data
}

\correspondingauthor{Yechi Zhang}
\email{yczhang@icrr.u-tokyo.ac.jp}

\author[0000-0003-3817-8739]{Yechi Zhang}
\affiliation{National Astronomical Observatory of Japan, 2-21-1 Osawa, Mitaka, Tokyo 181-8588, Japan}
\affiliation{Institute for Cosmic Ray Research, The University of Tokyo, 5-1-5 Kashiwanoha, Kashiwa, Chiba 277-8582, Japan}
\affiliation{Department of Astronomy, Graduate School of Science, the University of Tokyo, 7-3-1 Hongo, Bunkyo, Tokyo 113-0033, Japan}
\affiliation{Kavli Institute for the Physics and Mathematics of the Universe (Kavli IPMU, WPI), The University of Tokyo, 5-1-5 Kashiwanoha, Kashiwa, Chiba, 277-8583, Japan}

\author[0000-0002-1049-6658]{Masami Ouchi}
\affiliation{National Astronomical Observatory of Japan, 2-21-1 Osawa, Mitaka, Tokyo 181-8588, Japan}
\affiliation{Institute for Cosmic Ray Research, The University of Tokyo, 5-1-5 Kashiwanoha, Kashiwa, Chiba 277-8582, Japan}
\affiliation{Graduate University for Advanced Studies (SOKENDAI), 2-21-1 Osawa, Mitaka, Tokyo 181-8588, Japan}
\affiliation{Kavli Institute for the Physics and Mathematics of the Universe (Kavli IPMU, WPI), The University of Tokyo, 5-1-5 Kashiwanoha, Kashiwa, Chiba, 277-8583, Japan}

\author[0000-0003-2965-5070]{Kimihiko Nakajima}
\affiliation{National Astronomical Observatory of Japan, 2-21-1 Osawa, Mitaka, Tokyo 181-8588, Japan}

\author[0000-0002-6047-430X]{Yuichi Harikane}
\affiliation{Institute for Cosmic Ray Research, The University of Tokyo, 5-1-5 Kashiwanoha, Kashiwa, Chiba 277-8582, Japan}

\author[0000-0001-7730-8634]{Yuki Isobe}
\affiliation{Institute for Cosmic Ray Research, The University of Tokyo, 5-1-5 Kashiwanoha, Kashiwa, Chiba 277-8582, Japan}
\affiliation{Department of Physics, Graduate School of Science, The University of Tokyo, 7-3-1 Hongo, Bunkyo, Tokyo 113-0033, Japan}

\author[0000-0002-5768-8235]{Yi Xu}
\affiliation{Institute for Cosmic Ray Research, The University of Tokyo, 5-1-5 Kashiwanoha, Kashiwa, Chiba 277-8582, Japan}
\affiliation{Department of Astronomy, Graduate School of Science, the University of Tokyo, 7-3-1 Hongo, Bunkyo, Tokyo 113-0033, Japan}

\author[0000-0001-9011-7605]{Yoshiaki Ono}
\affiliation{Institute for Cosmic Ray Research, The University of Tokyo, 5-1-5 Kashiwanoha, Kashiwa, Chiba 277-8582, Japan}

\author[0009-0008-0167-5129]{Hiroya Umeda}
\affiliation{Institute for Cosmic Ray Research, The University of Tokyo, 5-1-5 Kashiwanoha, Kashiwa, Chiba 277-8582, Japan}
\affiliation{Department of Physics, Graduate School of Science, The University of Tokyo, 7-3-1 Hongo, Bunkyo, Tokyo 113-0033, Japan}

\begin{abstract}
We present statistics of $z\sim 6-9$ galaxy outflows indicated by spatially-extended gas emission and broad lines. With a total of 61 spectroscopically confirmed galaxies at $z\sim 6-9$ in the JWST CEERS, GLASS, and ERO data, we find four galaxies with [O{\sc iii}]+H$\beta$ ionized gas emission significantly extended beyond the kpc-scale stellar components on the basis of the emission line images constructed by the subtraction of NIRCam broadband (line on/off-band) images. By comparison with low-$z$ galaxies, the fraction of galaxies with the spatially extended gas, 4/18, at $z\sim 6-9$ is an order of magnitude higher than those at $z\sim 0-1$, which can be explained by events triggered by frequent major mergers at high redshift. We also investigate medium- and high-resolution NIRSpec spectra of 30 galaxies at $z\sim 6-9$, and identify five galaxies with broad ($140-800$~km~s$^{-1}$) lines in the [O{\sc iii}] forbidden line emission, suggestive of galaxy outflows. One galaxy at $z=6.38$ shows both the spatially-extended gas emission and the broad lines, while none of the galaxies with the spatially-extended gas emission or broad lines present a clear signature of AGN either in the line diagnostics or Type 1 AGN line broadening ($>1000$ km s$^{-1}$), which hint outflows mainly driven by stellar feedback. The existence of galaxies with/without spatially-extended gas emission or broad lines may be explained by different viewing angles towards outflows, or that these are galaxies in the early, late, post phases of galaxy outflows at high redshift, where the relatively large fractions of such galaxies indicate the longer-duration and/or more-frequent outflows at the early cosmic epoch.
%
\end{abstract}


\section{Introduction} \label{sec:intro}
Feedback has been known as the key to understanding galaxy formation and evolution. According to the $\Lambda$ cold dark matter ($\Lambda$CDM) model, galaxies build up their mass from the cool gas in the dark matter (DM) halos. Throughtout this process, different feedback mechanisms will affect the mass assembly of galaxies. For example, young, massive stars and/or active galactic niclei (AGNs) powered by the central supermassive black holes can produce feedback in the forms of ionizing radiation, stellar wind, and supernova explosion, heating the gas or expelling it from the star forming (SF) regions, thus limiting the SF activities. With the difficulty in fully reproducing these complicated physical processes solely with hydrodynamical simulations, observations, especially at the early epochs of galaxy formation, is in need to constrain and improve the simulation results. 

One of the direct phenomena of feedback is the gas outflow, which has been intensively studied with different techniques. From the Local Universe out to $z\sim6$, outflow is systematically investigated and identified with the metal absorption lines \citep[e.g.,][]{heckman00,heckman15,rupke05a,rupke05b,steidel10}, emission lines in the rest-frame optical \citep[e.g.,][]{cicone16,finley17,fs19,wylezalek20} and sub-mm wavelengths \citep[e.g.,][]{walter02,gallerani18,ginolfi20}, and absorption features in the background QSO continua \citep[e.g.,][]{bouche12,kac15,muzahid15}. These studies have found that outflow is positively correlated with stellar mass ($M_*$) and star formation rate (SFR) or galaxies, and are more frequently found in AGN.

With outflow evacuating gas from SF regions into the circum galactic medium (CGM) or even intergalactic medium (IGM), the gas sometimes is observed to be spatially extended beyond the galaxy continua. For example, \citet{harikane14} identified spatially extended [O{\sc ii}] emission around a $z=1.18$ SF galaxy. \citet{yuma13,yuma17} systematically investigated and identified the spatial extension of ionized gas traced by the rest-frame optical emission lines of [O{\sc ii}], H$\beta$, [O{\sc iii}], and H$\alpha$. With integral field unit (IFU) spectroscopic data, \citet{liu13a,liu13b} identified extended ionized gas nebulae around $z\sim0.5$ radio-quiet QSOs via [O{\sc ii}] emission lines. Similarly, \citet{rupke13} detected multi-phase outflows extending to kpc scale around ultraluminous infrared galaxies (ULIRGs). At the higher redshift of $z=4-6$, \citet{fujimoto20} identified the [C{\sc ii}]
metal emission lines in the rest-frame far-infrared (FIR) that extended to the CGM scale at $\sim10$~kpc in $30\%$ of their galaxies. With ALMA data, \citet{ginolfi20} also found the extended metal-enriched gas out to $\sim30$~kpc scale from stacking analysis. These studies have suggested that outflow driven by intensive SF activities or AGN may (partly) explain such spatially extended gas. However, the physical mechanism that triggers and connects the spatially extended gas and outflow activities still remain unclear.

With the launch of James Webb Space Telescope (JWST), it is now possible to directly observe and study galaxies at early epochs. For example, \citet{nakajima23} compiled the NIRSpec spectroscopy taken with three different JWST large programs and obtained a sample of 185 spectroscopically confirmed galaxies at $z=3.8-8.9$. These high-z galaxies open a new door to testing the evolution of the physical properties of galaxies, such as size, morphology, and metallicity. They also serve as a good sample from which systematic search for AGN and outflow activities can be performed.

In this paper, we search for ionized gas outflow traced by rest-frame optical [O{\sc iii}] $\lambda\lambda$4959, 5997 emission lines using JWST/NIRCam imaging and NIRSpec spectroscopy. From the \citet{nakajima23} sample, we identify spatial extension of [O{\sc iii}] emission around galaxies via the broadband excess technique, obtaining the [O{\sc iii}] emission line images by subtracting the stellar continua from the NIRCam broadband images that cover [O{\sc iii}] emission lines. We separately search for ongoing [O{\sc iii}] outflow by fitting the emission line profiles in the medium to high resolution NIRSpec spectroscopy. Section \ref{sec:data} summarizes the dataset and sample used in this study. In Section \ref{sec:blob} and \ref{sec:outflow} we select and analyze the galaxies with spatially extended [O{\sc iii}] emission and with ongoing [O{\sc iii}] outflow, respectively. Section \ref{sec:discussions} discusses the possible connection and origins of spatially extended [O{\sc iii}] emission and ongoing outflow, as well as the redshift evolution of their number fractions. Our findings are summarized in Section \ref{sec:summary}. Throughout the paper, we use AB magnitude and the cosmological parameters of \citet{planck18}.

\section{Data and Sample} \label{sec:data}
\subsection{JWST/NIRSpec Galaxies} \label{subsec:nirspec}
We use the JWST/NIRSpec spectroscopic dataset in \citet{nakajima23}. Here we briefly summarize the sample construction, and refer the readers to \citet{nakajima23} for the details of the JWST/NIRSpec observations and data reductions.

The spectroscopic dataset include Early Release Observations \citep[ERO;][]{p22} targeting the lensing cluster SMACS-0723, as well as two of the Early Release Sciences (ERS) observations, GLASS \citep[][]{treu22} that targets the lensing cluster of Abell-2744 and Cosmic Evolution Early Release Science \citep[CEERS;][]{bagley23,finkelstein23} that includes 12 pointings in the Extended Groth Strip (EGS) field. ERO data were taken with two medium resolution ($R\sim1000$) filter-grating pairs, F170LP-G235M and F290LP-G395M, covering the wavelength ranges of $1.7 - 3.1$ and $2.9 - 5.1$ $\mu$m, respectively. The total exposure time of the ERO data is 4.86 hours for each filter- grating pair.
GLASS data were taken with three high resolution ($R\sim2700$) filter-grating pairs, F100LP-G140H, F170LP-G235H, and F290LP-G395H, that cover 1.0-1.6, 1.7-3.1, and 2.9-5.1~$\mu$m, respectively. For each filter-grating pair used in GLASS observations, the total integration time was 4.9 hours. CEERS observations used both the Prism ($R\sim 100$) covering 0.6-5.3~$\mu m$ and medium resolution filter-grating pairs of F100LP-G140M, F170LP-G235M, and F290LP-G395M covering 1.0-1.6, 1.7-3.1, and 2.9-5.1~$\mu$m, respectively. The on-source integration time for CEERS is 0.86 hours. 

The aforementioned spectroscopic data were reduced in \citet{nakajima23} with the JWST pipeline version 1.8.5 and the Calibration Reference Data System (CRDS) files \texttt{jwst\_1027.pmap} and \texttt{jwst\_1028.pmap}. Additional processes were taken to improve the flux calibtaion, noise estimation, slit-loss correction, and stacking. From the composite spectra, \citet{nakajima23} visually identified the emission lines and measured the spectroscopic redshifts. In total, there are 185 galaxies with spectroscopically confirmed redshift at $z=3.8 - 8.9$. The $M_*$ and SFR of these galaxies were derived with SED fitting using NIRCam imaging (\ref{subsec:nircam}) and the total H$\beta$ luminosity, respectively. We refer the readers to \citet{nakajima23} for the details of the derivation.
\begin{figure}[ht!]
\begin{center}
\includegraphics[scale=0.33]{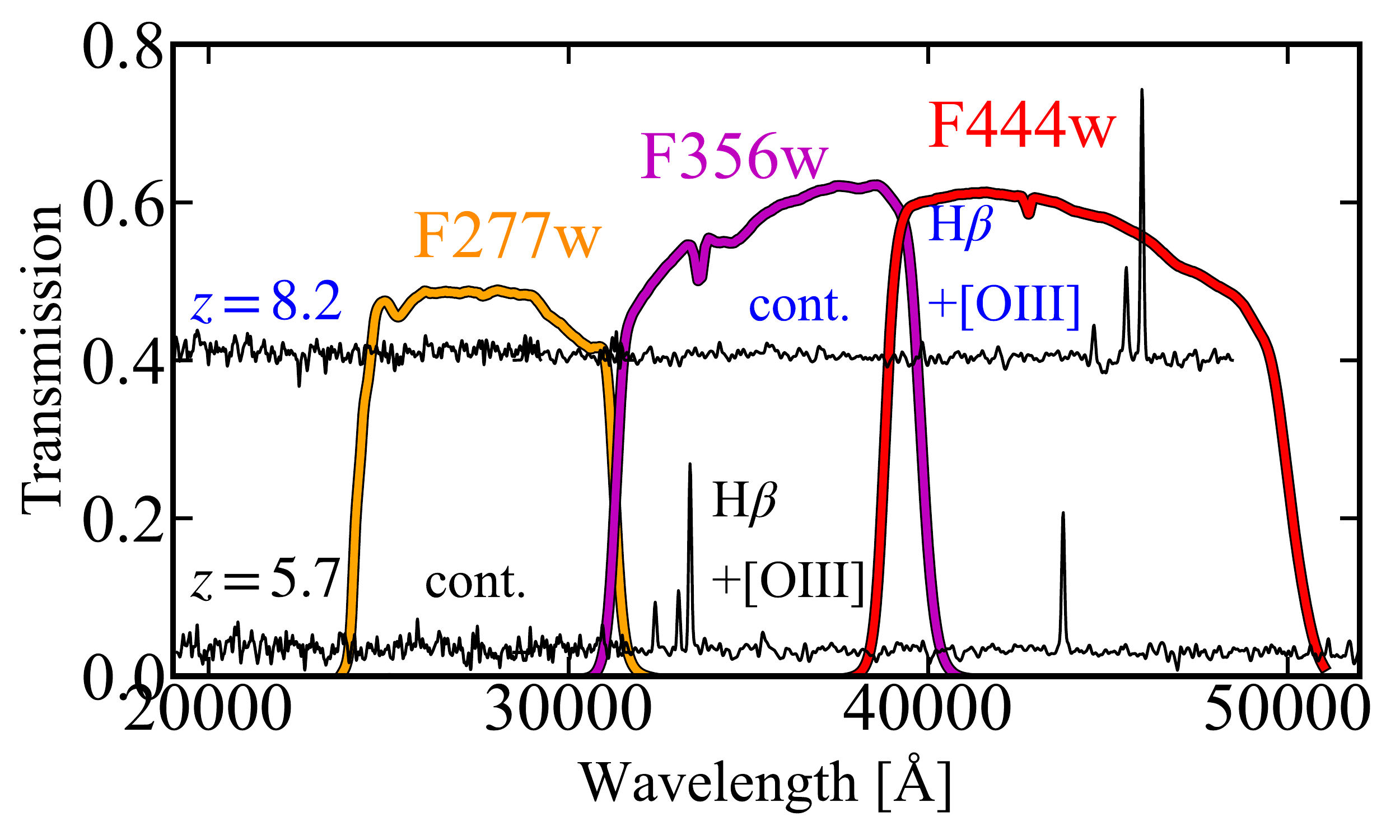}
\end{center}
\caption{Filter response curves of JWST/NIRCam F277w, F356w, and F444w broadband filters overplotted with example spectra of SF galaxies at $z\sim6$ and 8. At $z\sim6$, H$\beta$+[O{\sc iii}] emission lines are covered by F356w filter while F277 filter traces the optical continuum. At $z\sim8$, H$\beta$+[O{\sc iii}] emission lines are covered by F444w filter while F356 filter traces the optical continuum. \label{fig:filter}}
\end{figure}

\subsection{JWST/NIRCam Imaging} \label{subsec:nircam}
We utilize the JWST/NIRCam images covering the ERO, GLASS, and CEERS fields. ERO includes JWST/NIRCam data covering a $11.0$~arcmin$^2$ area around the lensing field of SMACS-0723 taken with the F090W, F150W, F200W, F277W, F356W, and F444W bands. CEERS program includes JWST/NIRCam images in 10 pointings with a total area of $\sim84
$~arcmin$^2$ taken with F115W, F150W, F200W, F277W, F356W, F410M, and F444W bands. For the GLASS NIRSpec field, we use the JWST/NIRCam data from the Ultradeep NIRSpec and NIRCam ObserVations before the Epoch of Reionization \citep[UNCOVER;][]{bezanson22} program. The UNCOVER team carried out NIRCam observations in a $6.8$~arcmin$^2$ area around the lensing field of Abell-2744 with seven bands, F090W, F115W, F150W, F200W, F277W, F356W, and F444W.

The ERO and CEERS NIRCam imaging data were reduced in \citet{harikane23a} and Harikane et al. (in preparation). They downloaded the raw data (\texttt{\_uncal.fits}) from the Mikulski Archive for Space Telescopes (MAST). Data reduction were performed with the JWST pipeline development version 1.6.3 and the CRDS file \texttt{jwst\_0995.pmap}. The pixel scale is set to 0."015/pixel. See \citet{harikane23a} and Harikane et al. (in preparation) for the details of data reduction. For the UNCOVER NIRCam data, we use the reduced images provided by the UNCOVER team \footnote{\url{https://jwst-uncover.github.io}} that have a pixel scale of 0."02/pixel in the short-wavelength bands (F090W, F115W, F150W) and 0."04/pixel in the long-wavelength bands (F200W, F277W, F356W, F444W) \citep{bezanson22}.
\begin{figure}[ht!]
\begin{center}
\includegraphics[scale=0.55]{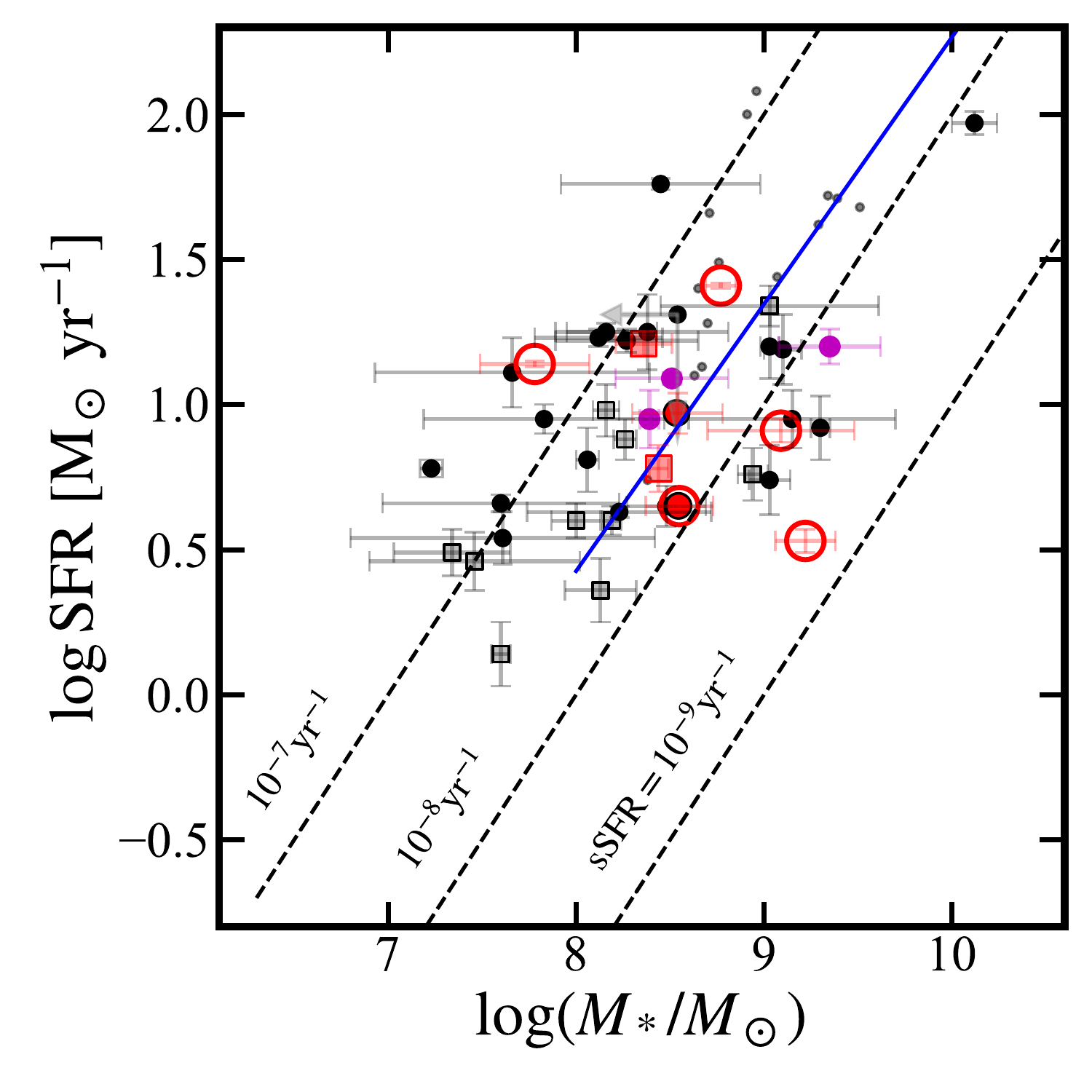}
\end{center}
\caption{$M_* - \mathrm{SFR}$ relation for galaxies at $z=5.4-8.9$ in the \citet{nakajima23} spectroscopic sample. Galaxies observed with medium- or high- resolution (Prism) spectra are labelled with circles (squares). The filled red symbols indicate the galaxies with spatially extended [O{\sc iii}] emission (see Section \ref{subsec:sample}). The red open circles shows the objects with ongoing [O{\sc iii}] outflow (see Section \ref{sec:outflow}). Objects with spatially offset or multi-component [O{\sc iii}] emission are shown in magenta square (see Section \ref{subsec:offset_multi}). The filled black symbols represents those objects without [O{\sc iii}] spatial extension or ongoing outflow. The small grey data points indicate objects without NIRCam data and are excluded from our analyses. Our sample galaxies are generally aligned with the SF main sequance at $z\sim6$ \citep[][blue line]{santini17}, and have specific SFR (sSFR) ranging from $10^{-9}-10^{-7}$~yr$^{-1}$.
\label{fig:ms_sfr}}
\end{figure}

\section{Spatially extended [O{\sc iii}] emission} \label{sec:blob}
\begin{figure}[ht!]
\begin{center}
\includegraphics[scale=0.55]{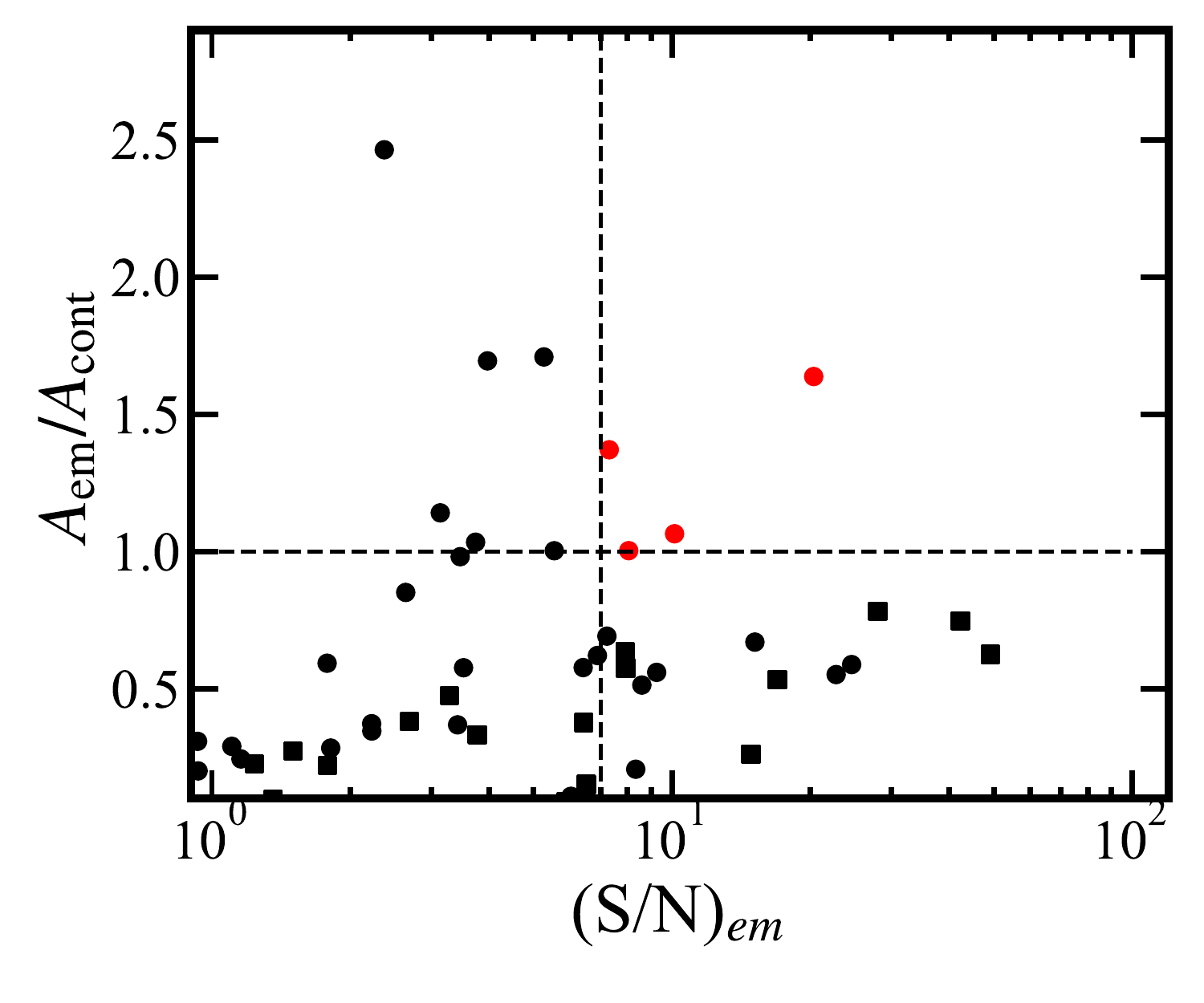}
\end{center}
\caption{$A_\mathrm{em}/A_\mathrm{cont}$ as a function of the $S/N$ ratio of the emission line images, where circles and squares inicate the $z=5-7$ and $z=7-9$ objects, respectively. The red symbols indicate our selected candicates that have spatially extended H$\beta$+[O{\sc iii}] emission. The black symbols shows galaxies without spatial extension of H$\beta$+[O{\sc iii}] emission. \label{fig:selection}}
\end{figure}
\subsection{Selection} \label{subsec:selection}
From the NIRCam images in Section \ref{subsec:nircam}, we construct the emission line images. For star-forming galaxies $z=5.4-7.0 (7.0-8.9)$, the H$\beta$ and [O{\sc iii}] emission lines fall into the wavelength range of F356W (F444W) band with the optical continua covered by the F277W (F356W) band. With the large H$\beta+$[O{\sc iii}] equivalent widths (EW) of the galaxies in \citet{nakajima23} sample that range from $95-8150$~\AA, the contribution of H$\beta+$[O{\sc iii}] emission lines to the F356W (F444W) is as large as 50\%. By subtracting the continuum images of F277W (F356W) band from those of F356W (F444W) band that contains the emission lines, we can obtain the emission line images without the optical continua. Before subtraction, we match the point spread function (PSF) of the continuum images to that of the F444W images in the same manner as \citet{harikane23a}. The typical FWHM of the PSF in F444W images is $\sim0.''16$. The typical $1\sigma$ noise level of the emission line images obtained is $\sim7\times10^{-17}$~erg~s$^{-1}$~cm$^{-2}$~arcsec$^{-2}$.

We find 37 (24) out of 52 (34) objects at $z=5.4-7.0 (7.0-8.9)$ in \citet{nakajima23} catalog (Section \ref{subsec:nirspec}) are covered by JWST/NIRCam footprint. These $37+24=61$ objects represents our sample. The $M_* - \mathrm{SFR}$ relation of the \citet{nakajima23} objects at $z=5.4-8.9$ with and without NIRCam coverage are presented in Figure \ref{fig:ms_sfr}. Our sample includes typical star-forming galaxies at $z\sim5-9$.

To select extended [O{\sc iii}] emission, we first require the emission line image to have $S/N >7$ measured within the $2\sigma$ isophotal areas. By this criteria we select 11 (7) out of the 37 (24) galaxies at $z=5.4-7.0 (7.0-8.9)$ with JWST imaging coverage. The EW of these 11+7=18 galaxies ranges between $803-6129$\AA.
We then compare the 2$\sigma$ isophotal areas of the continuum ($A_\mathrm{cont}$) and emission line ($A_\mathrm{em}$) images obtained with \texttt{SExtractor}. Previous studies often adopt a certain physical size of the isophotal area larger than which the object is classified as spatially extended \citep[e.g.,][]{yuma13}. However, the measured size of isophotal areas are subject to the surface brightness limits of the imaging data as well as cosmological dimming effect. In this work, we require spatially extended emission to have $A_\mathrm{em} > A_\mathrm{cont}$. Such a criteria ensures that the selected objects have larger spatial extension in the shallower emission line images compared with the continuum images. Figure \ref{fig:selection} shows our selection procedure, where the red data points show our extended [O{\sc iii}] emission candidates.

\subsection{Our Spatially Extended [O{\sc iii}] Emission Candidates} \label{subsec:sample}

\begin{figure*}[ht!]
\begin{center}
\includegraphics[scale=0.78]{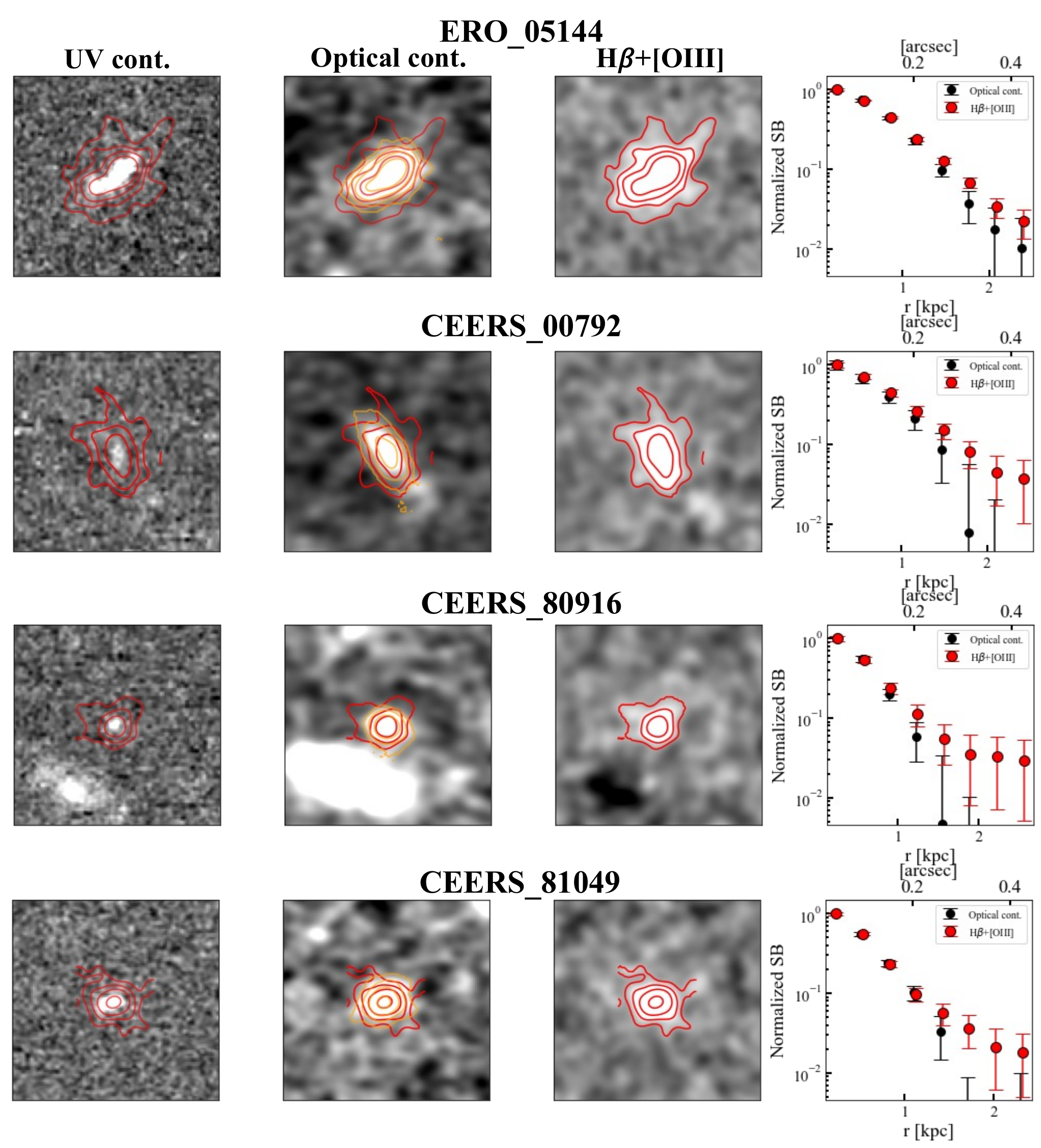}
\end{center}
\caption{Extended H$\beta$+[O{\sc iii}] emission candidates at $z=5.4-7$. From left to right: rest-frame UV (original psf) and optical (psf-matched) continuum images, H$\beta$+[O{\sc iii}] emission images (psf-matched), and radial profiles. On the images we overplot the emission line image and optical continuum image contours in red and orange, respectively. For the right panel, red (black) data points indicate the radial profiles of H$\beta$+[O{\sc iii}] (continuum) emission. \label{fig:z6b_img}} 
\end{figure*}

Based on the selection criteria mentioned above, we identify four galaxies with spatially extended [O{\sc iii}] emissions, all of which  
We present the IDs, redshift, and basic physical properties of the four extended [O{\sc iii}] emission candidates in Table \ref{tab:blob}. At $z=5.4-7.0$, there are four out of 37 galaxies that have extended [O{\sc iii}] emissions. One of the four objects, ERO\_05144, is a lensed galaxy with a magnification factor $\mu = 3.18$. In Figure \ref{fig:z6b_img}, we present the rest-frame ultraviolet (UV), rest-frame optical continuum, and the [O{\sc iii}] emission images, together with the radial profiles derived with respect to the center positions in the rest-frame optical continuum images. In the [O{\sc iii}] emission line images, all of these four objects have [O{\sc iii}] emission extended beyond the stellar continua, while the radial profiles also confirm such extensions beyond $\sim1.5$~kpc. 

We further characterize the radial surface brightness profiles of these four objects by two-dimensional fitting with \texttt{GALFIT}. Here we assume the radial surface brightness profiles of both the optical continuum and [O{\sc iii}]+H$\beta$ emission are described by a S\'ersic profile with S\'ersic index $n=1$ (i.e., exponential profile), which have been adopted in the measurement of spatially extended gas \citep[e.g.,][]{steidel11,farina19,fujimoto19}. For each object, we obtain the PSF by selecting and stacking bright stars in the same field. As listed in Table \ref{tab:blob}, we confirm that all of the four objects have effective radii of emission line profile ($r_{e,\mathrm{em}}$) larger than those of the optical continuum profile ($r_{e,\mathrm{cont}}$).

\begin{deluxetable*}{lccccccccc}[hbtp!]
\tablecaption{Galaxy and imaging properties of spatially extended H$\beta$+[O{\sc iii}] emission candidates \label{tab:blob}}
\tablewidth{0pt}
\tablehead{
\colhead{ID} &
\colhead{$z_\mathrm{spec}$} &
\colhead{$\log M_*$\tablenotemark{*}} &
\colhead{$\log$ SFR\tablenotemark{*}} &
\colhead{$A_\mathrm{cont}$} &
\colhead{$A_\mathrm{em}$} &
\colhead{$r_e\mathrm{,cont}$} &
\colhead{$r_e\mathrm{,em}$} &
\colhead{Spec\_flag\tablenotemark{\textdagger}} &
\colhead{Broad [O{\sc iii}]} \\
 & & ($M_\odot$) & ($M_\odot$~yr$^{-1}$) & (arcsec$^2$) & (arcsec$^2$) & (arcsec) & (arcsec) & & 
}
\startdata
\multicolumn{10}{c}{$z=5.4-7$} \\ 
ERO\_05144 & 6.378 & $8.55^{+0.18}_{-1.03}$ & $0.65^{+0.01}_{-0.01}$ & $0.229$ & $0.374$ & $0.079\pm0.001$ & $0.084\pm0.001$ & M & True  \\ 
CEERS\_00792 & 6.257 & $8.54^{+0.24}_{-0.66}$ & $0.97^{+0.07}_{-0.08}$ & $0.194$ & $0.207$ & $0.146\pm0.007$ & $0.182\pm0.002$ & M & False \\
CEERS\_80916 & 5.674 & $8.44^{+0.05}_{-0.23}$ & $0.78^{+0.08}_{-0.10}$ & $0.098$ & $0.134$ & $0.028\pm0.002$ & $0.037\pm0.002$ & P & - \\
CEERS\_81049 & 6.738 & $8.36^{+0.15}_{-0.05}$ & $1.21^{+0.04}_{-0.05}$ & $0.137$ & $0.138$ & $0.050\pm0.001$ & $0.066\pm0.002$ & P & - \\
\hline
\enddata
\tablenotetext{*}{Obtained from \citet{nakajima23}}
\tablenotetext{$\textdagger$}{P: Prism; M: Medium-resolution filter-grating pair}
\end{deluxetable*}

\subsection{Spatially Offset and Multi-Component Emission} \label{subsec:offset_multi}
\begin{figure}[ht!]
\begin{center}
\includegraphics[scale=0.5]{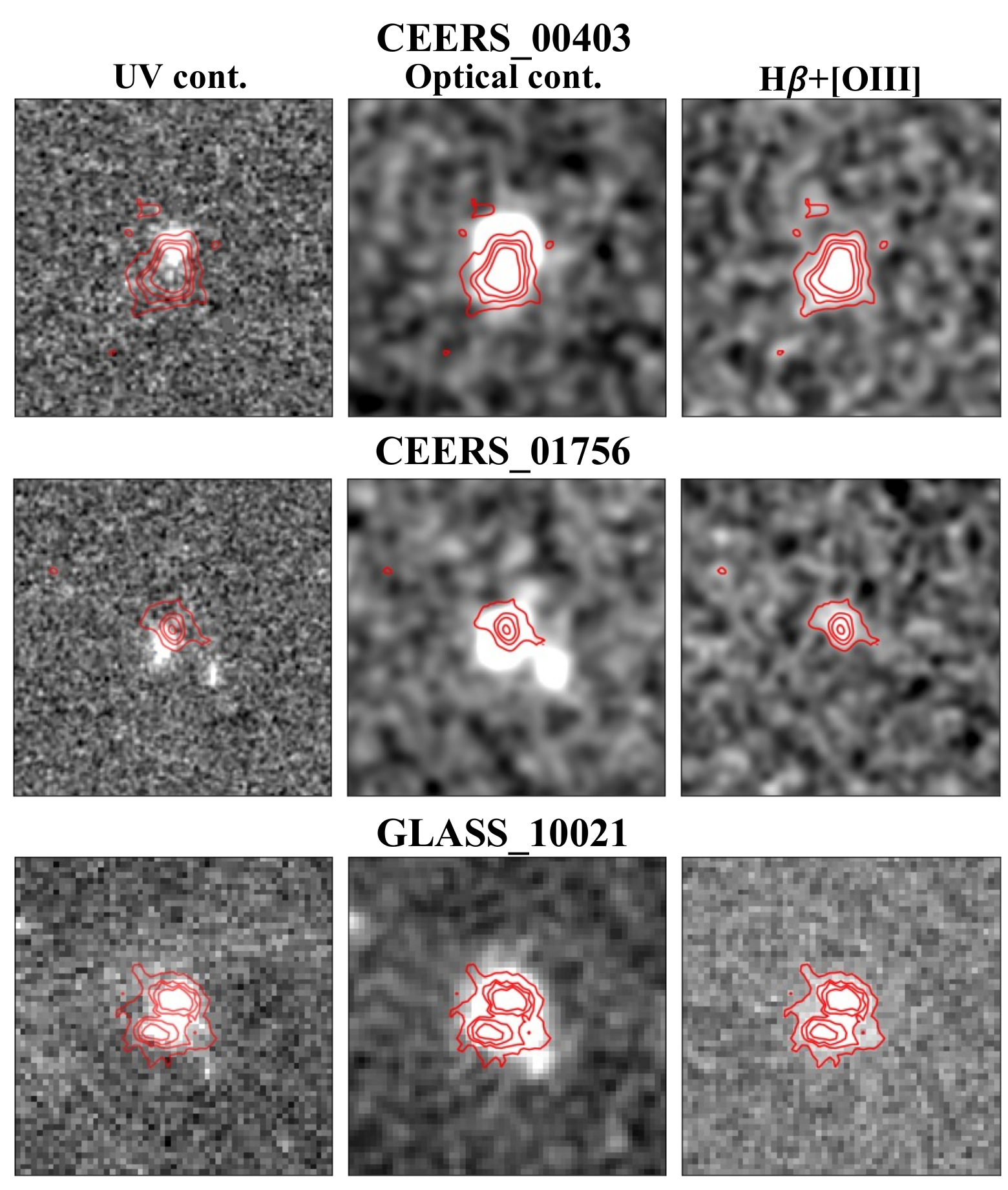}
\end{center}
\caption{Same as Figure \ref{fig:z6b_img}, but for the three objects with spatially offset or multi-component [O{\sc iii}] emission. \label{fig:z6o_img}}
\end{figure}

Through examination of emission line images and radial profiles, we also serendipitously identify three objects that, although do not satisfy the criteria of extended [O{\sc iii}] emissions, have [O{\sc iii}] emissions that are spatially offset from their stellar continua or have multiple components. Figure \ref{fig:z6o_img} shows the images of these three objects. For CEERS\_00403 and CEERS\_01756, 
We find that the [O{\sc iii}] emission is offset from the rest-frame UV and optical continuum, which may indicate the regions inside the galaxies where intense SF activities are ongoing. GLASS\_10021 is a lensed galaxy at $z=7.29$ with a lensing factor of $\mu=1.72$. From Figure \ref{fig:z6o_img}, we tentatively observe two components in the emission line image, although these two components are unresolved. The peak positions of the two components are separated by $0.''26$. Because such a multi-component feature is not observed in the stellar continuum, it is unlikely that the multi-component [Oiii] emission is due to lensing effect.
\begin{figure*}[ht!]
\begin{center}
\includegraphics[scale=0.73]{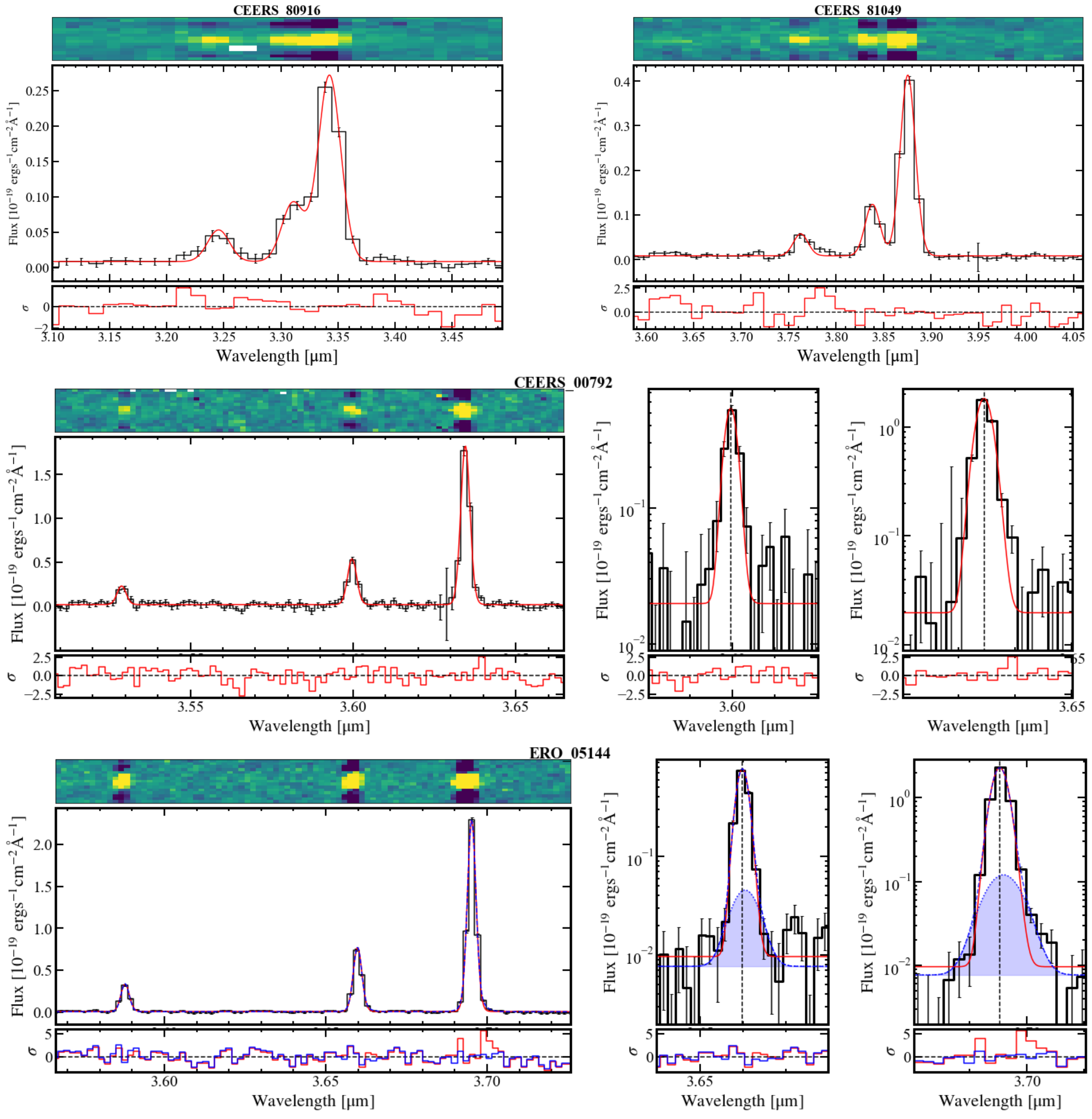}
\end{center}
\caption{1D and 2D NIRCam spectra of the four objects with spatially extended [O{\sc iii}] emission, cut out at the wavelength range around observed H$\beta$ and [O{\sc iii}] emission lines. The top row shows the two objects with Prism spectra. The next three row show the three objects with medium- or high-resolution spectra, while the middle and right panels correspond to the zoom-in 1D spectral region of [O{\sc iii}]$\lambda4959$ and $\lambda5007$, respectively, with the log scale in flux densities. The red curves indicate the best-fit no-outflow model (see texts), while the blue curves and shaded regions highlight the outflow components when available. The residuals of best-fit models are shown at the bottom of each spectrum. \label{fig:bl_spec}}
\end{figure*}

\subsection{Spectroscopy of Extended [O{\sc iii}] Emission} \label{subsec:blob_spec}
We examine the existence of broad line [O{\sc iii}] outflows in NIRSpec spectra of the four objects with spatially extended [O{\sc iii}] emission (Section \ref{subsec:sample}). In Figure \ref{fig:bl_spec}, we show the H$\beta$ and [O{\sc iii}] emission lines of these five objects. Two of the objects, CEERS\_80916 and CEERS\_81049, were taken with Prism that has a low spectral resolution of $R\sim100$, while the remaining three objects were taken with medium resolution filter-grating pairs. 

We conduct emission line fitting on H$\beta$ and [O{\sc iii}]$\lambda\lambda$4959, 5007 simultaneously, assuming a model with a flat continuum and one Gaussian profile for each of the three emission lines (hereafter ``no-outflow model"). For the Gaussian components, we tie the rest-frame central wavelength of H$\beta$, [O{\sc iii}]$\lambda$4959, and [O{\sc iii}]$\lambda$5007 to 4862.68\AA, 4960.30\AA, and 5008.24\AA, respectively, and assume H$\beta$ and [O{\sc iii}]$\lambda\lambda$4959, 5007 have the same line widths (FWHM$_\mathrm{n}$) in the unit of km~s$^{-1}$. 
We then convolve our model with the line spread function (LSF) of \citet{isobe23} and fit it to the observed data. 
We obtain the best-fit model with Monte Carlo simulations, making 500 mock spectra by adding random noise to the observed flux density of each spectral pixel. The random noise is generated following a Gaussian distribution, whose standard deviation is defined by the $1\sigma$ uncertainty in the observed spectra. The best-fit parameters and their errors are obtained with the mean and standard deviation of the distribution of the output parameters through 500 iterations. 

As shown in Figure \ref{fig:bl_spec} and Table \ref{tab:blob}, three out of the four objects have H$\beta$ and [O{\sc iii}]$\lambda\lambda$4959, 5007 emission lines that can be well-fitted with single, narrow Gaussian profiles without large residuals. For the two objects observed with Prism, CEERS\_80916 and CEERS\_81049, the spectral resolution is too low to resolve the outflow components. No conclusions can be made on the existence of ongoing outflow in these two objects. For CEERS\_00792 that was observed with medium-resolution filter-grating pairs, we find no outflow components in their spectra. The remaining object, ERO\_05144, has a $4-5\sigma$ residual around the [O{\sc iii}]$\lambda$5007 emission line, indicating the possible existence of ongoing outflow. 

We fit the H$\beta$ and [O{\sc iii}]$\lambda\lambda$4959, 5007 emission lines of ERO\_05144 again, adding a broad Gaussian component to the [O{\sc iii}]$\lambda$4959 and [O{\sc iii}]$\lambda$5007 emission line, respectively (hereafter ``outflow model"). We tie the velocity shifts ($\Delta v$, in km~s$^{-1}$) and the line widths (FWHM$_\mathrm{b}$, in km~s$^{-1}$) of these two Gaussian components. For the broad component, we require FWHM$_\mathrm{b} > $FWHM$_\mathrm{n}$ without setting a specific threshold value for FWHM$_\mathrm{b}$ \citep[e.g.,][]{swinbank19}. We also fix the flux ratio ($fratio$) of the broad and narrow components for [O{\sc iii}]$\lambda$4959 and [O{\sc iii}]$\lambda$5007 emission lines. 
We then repeat the fitting process, obtaining the best-fit model (blue curves in Figure \ref{fig:bl_spec}) and parameters (Table \ref{tab:outflow}). It is clearly seen that the outflow model is preferred over the no-outflow one for the spectra of ERO\_05144. We compare the difference in Bayesian information criterion \citep[$\Delta$BIC;][]{schwarz78} for the two models, which is given by $\Delta\mathrm{BIC} = \chi^2_2 - \chi^2_1 + (k_2-k_1)\log N$. Here $\chi^2_1$($\chi^2_2$) and $k_1$($k_2$) refers to the $\chi^2$ and number of free parameters for no-outflow (outflow) model, respectively. For ERO\_05144, we obtain $\Delta$BIC $=46.4$, suggesting that the outflow model better describes the observed data. 

We also examine the NIRSpec data of the three objects that have offset or multi-component [O{\sc iii}] spatial distribution. One of the objects, GLASS\_10021, has a FWHM$_\mathrm{b}$ that is almost identical to FWHM$_\mathrm{n}$ with two components separated by 76~km~s$^{-1}$. This objects, mentioned previously in Section \ref{subsec:offset_multi}, also has two components in the [O{\sc iii}] emission line image. It is possible that we are observing the rotation disk of this object. For the other two objects, CEERS\_00403 and CEERS\_01756, the emission lines are well fitted with the no-outflow model, indicating that the offset spatial distribution of [O{\sc iii}] is unlikely driven by stellar outflow. 

\section{Broad [O{\sc iii}] Emission in NIRSpec Spectroscopy} \label{sec:outflow}
\begin{deluxetable*}{lcccccccccc}[htbp!]
\tablecaption{Galaxy and emission line properties of [O{\sc iii}] outflows \label{tab:outflow}}
\tablewidth{0pt}
\tablehead{
\colhead{ID} & 
\colhead{$z_\mathrm{spec}$} &
\colhead{$\log M_*$\tablenotemark{*}} &
\colhead{$\log$ SFR\tablenotemark{*}} &
\colhead{$v_\mathrm{esc}$} &
\colhead{FWHM$_\mathrm{b}$} &
\colhead{$\Delta v$} &
\colhead{$v_\mathrm{max}$} &
\colhead{Flux$_\mathrm{b}$\tablenotemark{\textdagger}} &
\colhead{(S/N)$_\mathrm{b}$} &
\colhead{$\Delta$BIC} \\
 & & ($M_\odot$) & ($M_\odot$~yr$^{-1}$) & (km~s$^{-1}$) & (km~s$^{-1}$) & (km~s$^{-1}$) & (km~s$^{-1}$) & ($10^{-17}$~erg~s$^{-1}$) & &
}
\startdata
\multicolumn{11}{c}{$z=5.4-7.0$} \\ 
ERO\_05144 
& 6.378 & $8.55^{+0.18}_{-1.03}$ & $0.65^{+0.01}_{-0.01}$ & $433.92^{+59.95}_{-343.04}$ & $568.16_{-66.25}^{+75.10}$ & $48.76_{-21.77}^{+30.69}$ & $332.84_{-48.40}^{+58.10}$ & $8.62_{-1.57}^{+2.04}$ & 4.9 & 46.4  \\ 
GLASS\_50038 
& 5.772 & $9.09^{+0.39}_{-0.09}$& $0.91^{+0.04}_{-0.05}$ & $533.43^{+32.75}_{-4.09}$ & $137.39_{-27.70}^{+33.22}$ & $141.50_{-20.53}^{+12.60}$ & $210.19_{-9.39}^{+9.03}$ & $13.24_{-2.07}^{+3.97}$ & 3.6 & 263.3 \\
GLASS\_110000 
& 5.763 & $9.22^{+0.16}_{-0.27}$ & $0.53^{+0.04}_{-0.05}$ & $572.00^{+127.32}_{-13.17}$ & $193.97_{-8.93}^{+7.81}$ & $-50.35_{-5.21}^{+5.08}$ & $147.33_{-4.79}^{+5.35}$ & $29.33_{-1.47}^{+1.34}$ & 15.7 & 148.5 \\
\hline
\multicolumn{11}{c}{$z=7.0-8.9$} \\ 
ERO\_06355 
& 7.665 & $8.77^{+0.08}_{-0.01}$ & $1.41^{+0.01}_{-0.01}$ & $524.06^{+64.36}_{-108.60}$ & $802.85_{-117.08}^{+166.67}$ & $59.08_{-16.97}^{+19.78}$ & $460.50_{-69.33}^{+85.26}$ & $23.65_{-2.86}^{+4.34}$  & 5.3 & 159.5 \\
ERO\_10612 
& 7.660 & $7.78^{+0.29}_{-0.03}$ & $1.14^{+0.01}_{-0.02}$ & $333.54^{+99.84}_{-23.04}$ & $481.83_{-61.43}^{+129.50}$ & $4.20_{-12.90}^{+16.08}$ & $245.11_{-34.85}^{+72.58}$ & $18.79_{-5.70}^{+6.79}$ & 3.2 & 36.0 \\
 \hline
\enddata
\tablenotetext{*}{Obtained from \citet{nakajima23}}
\tablenotetext{$\textdagger$}{Flux of the broad component of [O{\sc iii}]$\lambda$5007.}
\end{deluxetable*}
\begin{figure*}[ht!]
\begin{center}
\includegraphics[scale=0.86]{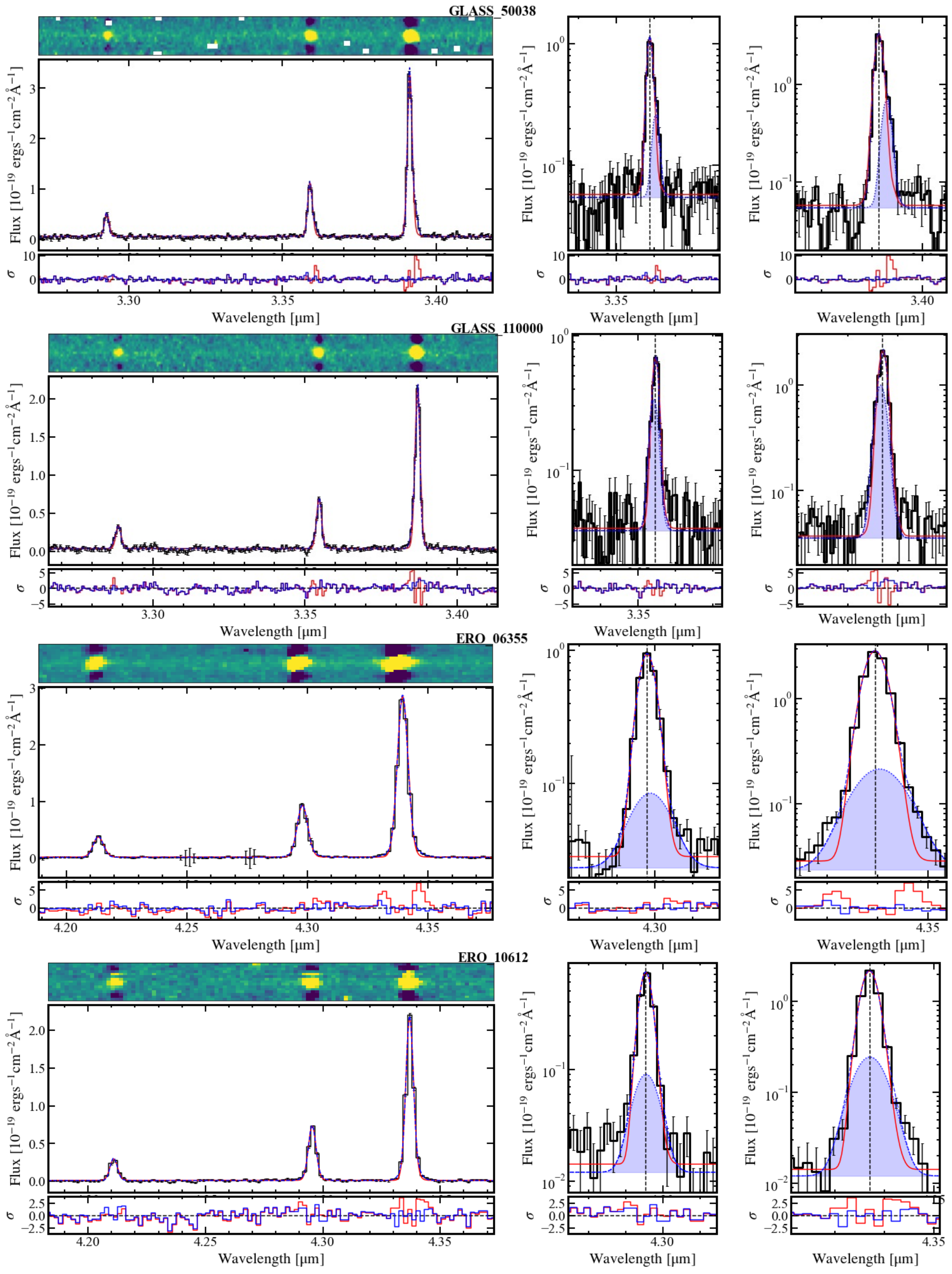}
\end{center}
\caption{Same as Figure \ref{fig:bl_spec}, but for the remaining four galaxies with ongoing outflow apart from ERO\_05144. \label{fig:out_spec}}
\end{figure*}
%
%

%
In this section we investigate broad line [O{\sc iii}] outflow with medium and high resolution NIRSpec data. With the difficulty in resolving outflow components with the low resolution of Prism spectroscopy ($R\sim100$), we focus on objects at $z=5.4-8.9$ taken with the medium and high resolution filter-grating pairs in \citet{nakajima23} sample, giving a total number of 30 galaxies.

For each of the 30 galaxies, we conduct emission line fitting on H$\beta$ and [O{\sc iii}]$\lambda\lambda$4959, 5007 as mentioned in Section \ref{subsec:blob_spec}, with no-outflow and outflow models. We select objects with ongoing outflow based on the following two criteria:
\begin{itemize}
    \item The observed data is preferred by the outflow model over the no-outflow one, which is quantified by $\Delta\mathrm{BIC} > 0$. 
    \item The best-fit outflow components of [O{\sc iii}]$\lambda\lambda$4959, 5007 have a signal to noise ratio (S/N) greater than 3.
\end{itemize}
With the two criteria mentioned above, we obtain five out of 30 objects that have [O{\sc iii}] outflow components whose emission line properties are listed in Table \ref{tab:outflow}. As shown in Figure \ref{fig:ms_sfr}, these five galaxies do not have significant dependence on either $M_*$ or SFR, which is similar to spatially extended objects (Section \ref{subsec:sample}). Out of five galaxies with ongoing outflow, only one object, ERO\_05144, shows spatially extended [O{\sc iii}] emission. 

\begin{figure}[ht!]
\begin{center}
\includegraphics[scale=0.49]{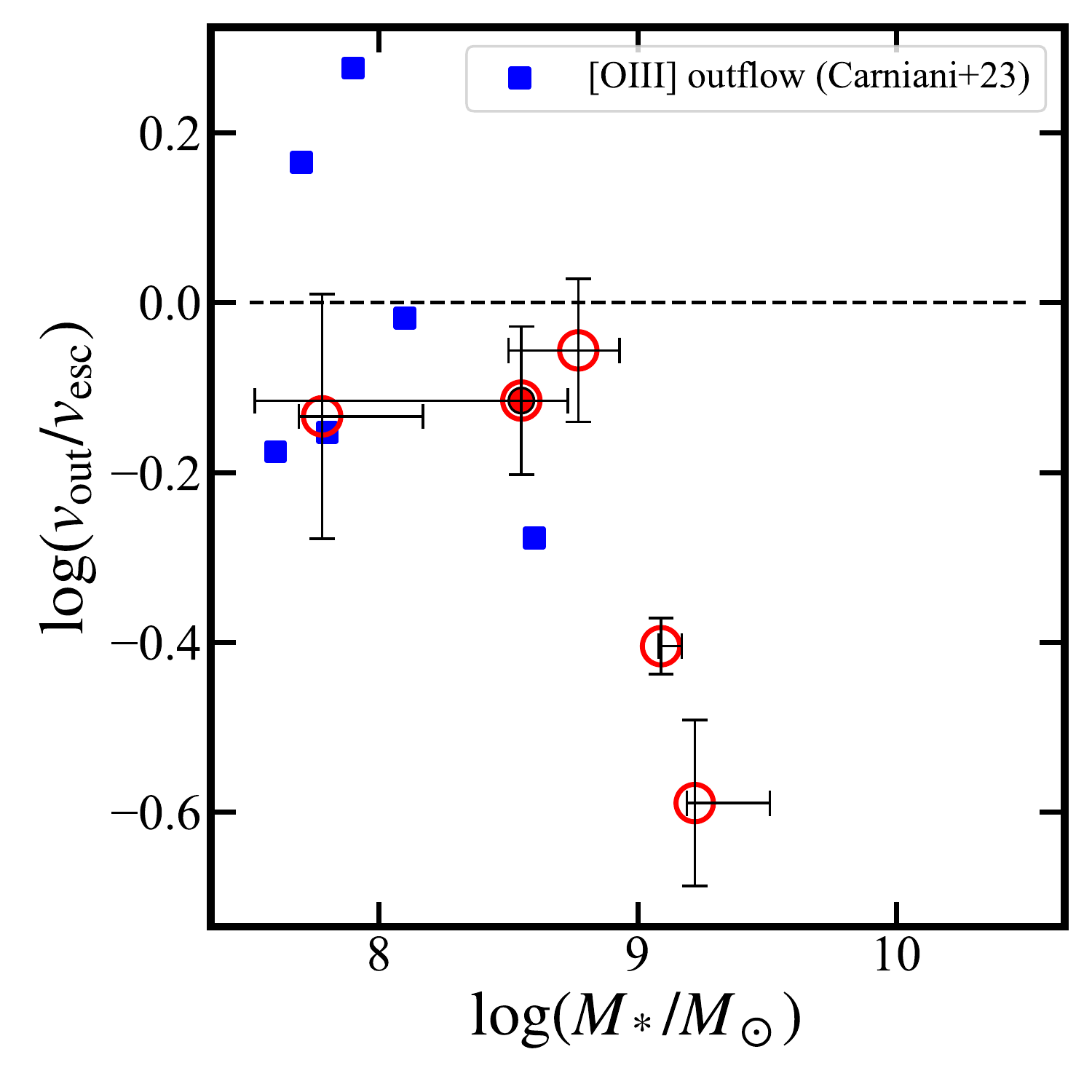}
\end{center}
\caption{$v_\mathrm{max}/v_\mathrm{esc}$ as a function of $M_*$. Symbols are identical to Figure \ref{fig:ms_sfr}, except that the blue squares denote the [O {\sc iii}] outflows at $z\sim4-7$ from \citet{carniani23}.  \label{fig:vesc}}
\end{figure}

From the FWHM measurement of the narrow and broad [O{\sc iii}] components, we calculated the maximum velocity that is defined as the line-of-sight velocity in the $\Delta v$ direction:
\begin{equation} \label{eq:vmax}
    v_\mathrm{max} = \mathrm{FWHM_b} + |\Delta v|/2.
\end{equation}
We compare $v_\mathrm{max}$ with the escape velocity ($v_\mathrm{esc}$) to estimate whether or not the ionized gas outflow can escape from the host dark matter (DM) halo. To obtain $v_\mathrm{esc}$, we first estimate the circular velocity ($v_\mathrm{cir}$) that is dependent on the DM halo mass ($M_\mathrm{h}$) with the following relations \citep{mo02} :
\begin{equation} \label{eq:vcir}
    v_\mathrm{cir} = \left(\frac{GM_\mathrm{h}}{r_\mathrm{h}}\right)^{1/2},
\end{equation}
\begin{equation} \label{eq:rh}
    r_\mathrm{h} = \left(\frac{GM_\mathrm{h}}{100\Omega_\mathrm{m}H_0^2}\right)^{1/3}(1+z)^{-1},
\end{equation}
where $M_\mathrm{h}$ and its scatter is obtained from the $M_*-M_\mathrm{h}$ relation at $z\sim7$ presented in \citet{behroozi19}. We convert the derived $v_\mathrm{cir}$ to $v_\mathrm{esc}$ with $v_\mathrm{esc} = 3v_\mathrm{cir}$ as adopted in previous studies \citep[e.g.,][]{xu22}. The errors in $v_\mathrm{esc}$ are calculated with the quadrature sum of the measurement errors propagated from $M_*$ and the typical 0.2~dex systematic errors of the $M_*-M_\mathrm{h}$ relation. In Figure \ref{fig:vesc}, we present the relation between $v_\mathrm{max}/v_\mathrm{esc}$ and $M_*$ for our galaxies with ongoing outflow. All of the five galaxies have $v_\mathrm{max}/v_\mathrm{esc} < 1$, indicating that the ongoing outflow wind is unlikely to escape from the DM halos. We compare our results with \citet{carniani23}, who also identified [O {\sc iii}] outflows in $z\sim4-7$ galaxies from JADES spectroscopic data. We adopt the emission line measurements and $M_*$ estimation from Table 2 and 3 of \citet{carniani23}, respectively, and derive $v_{\rm max}/v_{\rm esc}$ in the aforementioned manner. As shown in Figure \ref{fig:vesc}, our results are generally consistent with the \citet{carniani23} sample, although $v_{\rm max}/v_{\rm esc}$ may anti-correlate with $M_*$ despite small statistics.

\section{Discussions} \label{sec:discussions}
\begin{figure}[ht!]
\begin{center}
\includegraphics[scale=0.5]{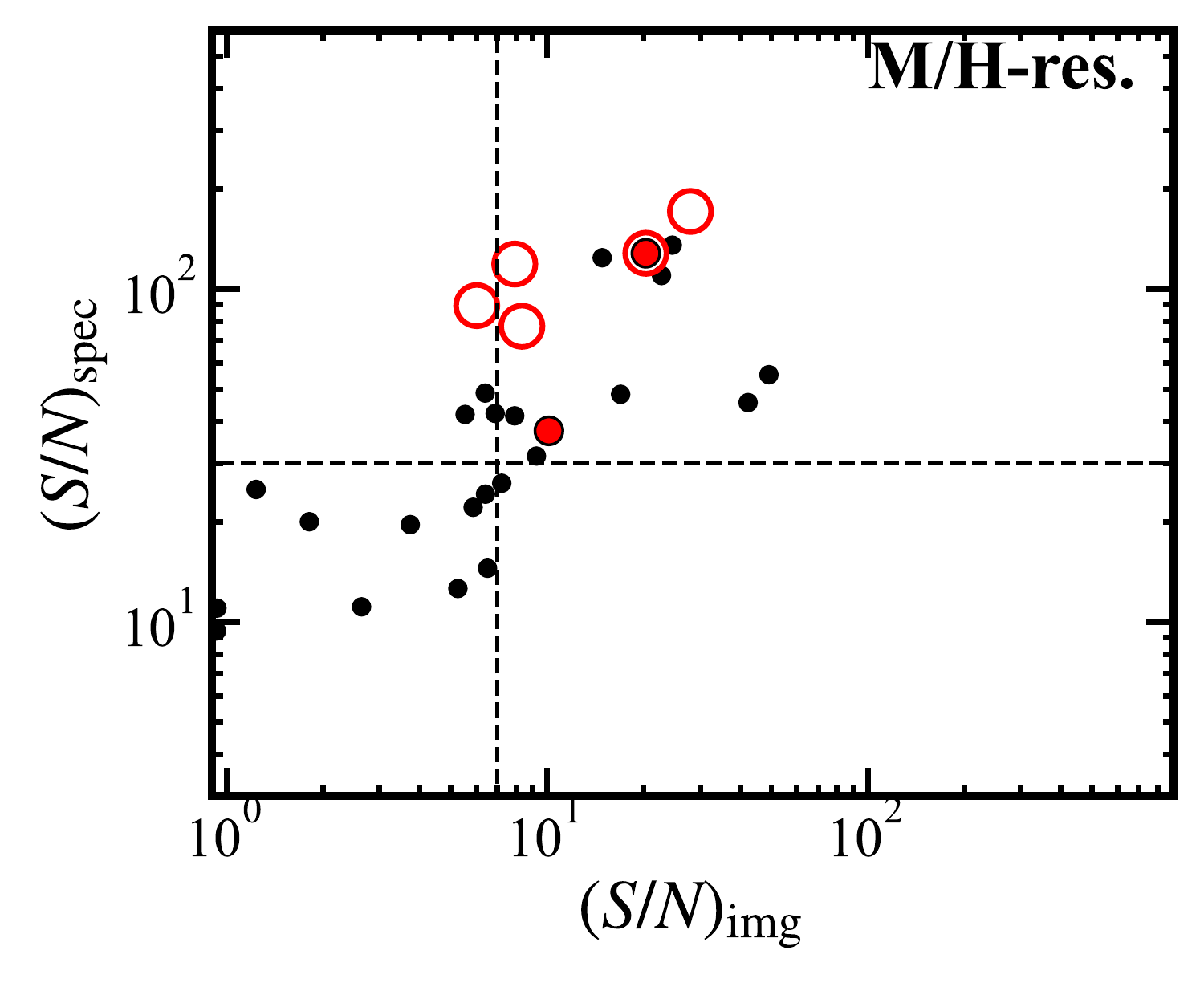}
\end{center}
\caption{The $S/N$ cuts for imaging and spectroscopic data of  the 30 objects with medium/high-resolution spectra, where the symbols are identical to Figure \ref{fig:ms_sfr}. \label{fig:sncut}}
\end{figure}
\subsection{Number Fractions of [O{\sc iii}] Spatial Extension and Ongoing Outflow}\label{subsec:number_frac}

As presented in Section \ref{subsec:sample} and \ref{sec:outflow}, out of the 61 galaxies with NIRCam imaging coverage at $z=5.4-8.9$ from \citet{nakajima23} sample, 31(30) galaxies were observed with the prism (medium/high-resolution filter-grating pairs). We have selected four galaxies with spatially extended [O{\sc iii}] emission from the 61 galaxies with imaging coverage, and five galaxies with broad [O{\sc iii}] emission lines from the 30 galaxies with medium/high-resolution spectra. To ensure that the non-detections of spatial extension and broad Gaussian components in [O{\sc iii}] emission lines are not due to the detection limits, we limit our discussions to the objects with high $S/N$. Specifically, we require $S/N$ of the NIRCam emission line image ($S/N_\mathrm{img}$) to be greater than 7 and the S/N of NIRSpec [O{\sc iii}] emission lines ($S/N_\mathrm{spec}$) to be greater than 30 for any potential detections of spatial extension and broad line components, respectively. With these criteria, we find 18/61 galaxies have  $S/N_\mathrm{img}>7$ in NIRCam emission line images, and 17/30 galaxies have $S/N_\mathrm{spec}>30$ in medium/high-resolution NIRSpec [O{\sc iii}] detection (Figure \ref{fig:sncut}). From these S/N cuts, we obtain the number fraction of spatially extended objects ($f_\mathrm{ext}$) to be $4/18=22.2\%$ (36.4\% and $<14.3\%$ at $z=5.7-7.0$ and $7.0-8.9$, respectively), and the number fraction of objects with broad [O{\sc iii}] emission lines ($f_\mathrm{broad}$) to be $5/17=29.4\%$ (30.0\% and 28.6\% at $z=5.7-7.0$ and $7.0-8.9$, respectively).

\begin{figure}[ht!]
\begin{center}
\includegraphics[scale=0.38]{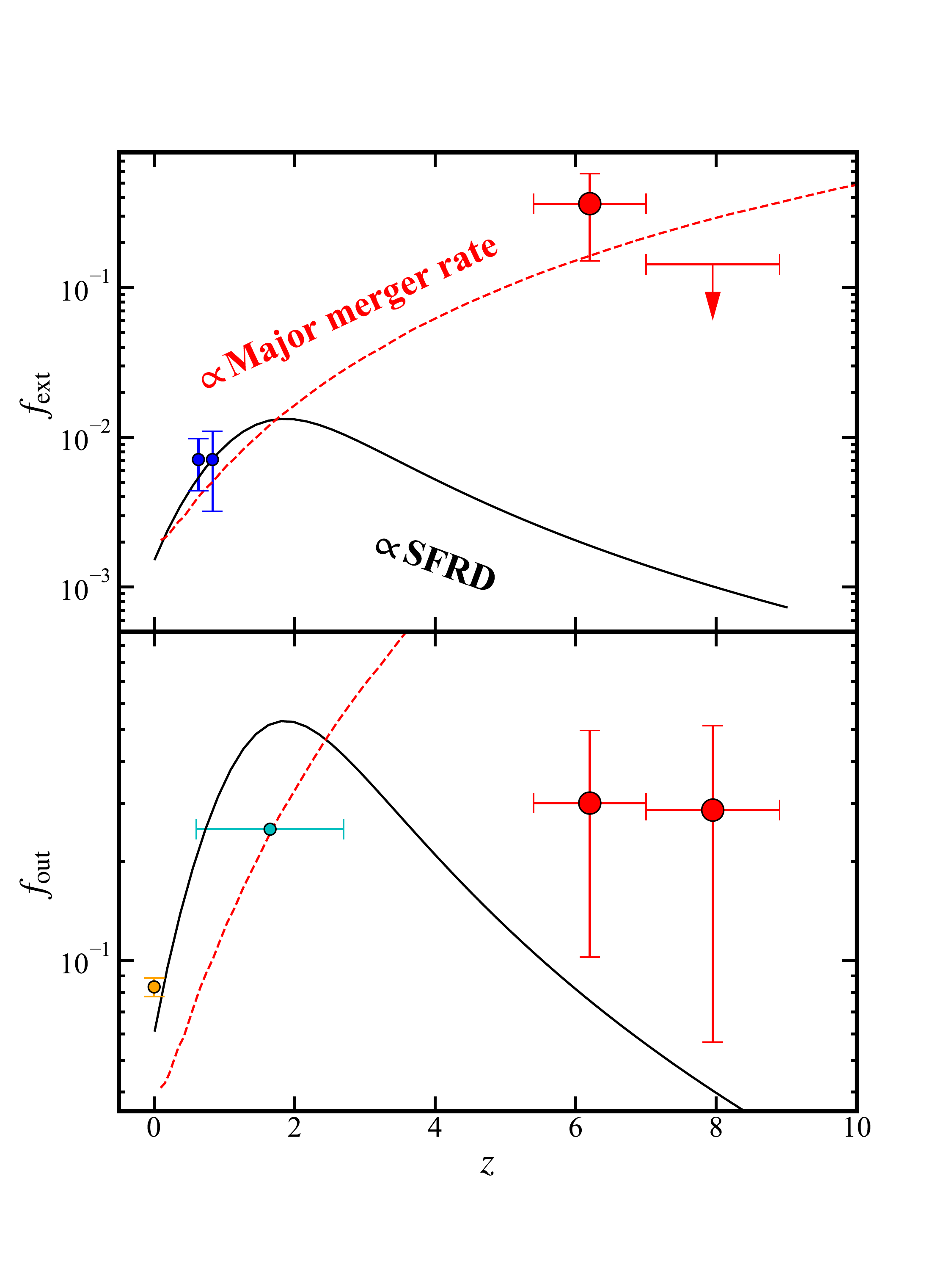}
\end{center}
\caption{Top: redshift evolution of spatially extended [O{\sc iii}] emission fraction. Our results are shown in red circles. The measurements at low redshift by \citet{yuma17} are indicated with blue circles. We also scale and overplot the redshift evolution of cosmic SFRD \citep[black solid curve;][]{madau14} and halo merger rate \citep[red dashed curve;][]{fakhouri10}. Bottom: redshift evolution of ongoing outflow fraction. Same as the top panel, our results are denoted with red circles. The yellow and magenta circles indicate the number fraction derived by \citet{wylezalek20} and \citet{fs19} at $z\sim0$ and $z=0.6-2.7$, respectively. The meanings of black solid curve and red dashed curve are identical to the top panel. \label{fig:bfrac}}
\end{figure}

In the top panel of Figure \ref{fig:bfrac}, we compare these number fractions with those obtained at $z<1$ by \citet{yuma17}, who identified spatially extended [O{\sc iii}] emission from $M_*\sim10^8-10^{10}~M_\odot$ galaxies, which is similar to our sample, using the narrowband excess technique. With a surface brightness limit of $1.2\times10^{-18}$~erg~s$^{-1}$~cm$^{-2}$~arcsec$^{-2}$, they reported $f_\mathrm{ext}$ values of 7/985 and 13/930 at $z=0.63$ and 0.83, respectively. Accounting for the surface brightness dimming, this limit corresponds to objects with $2\times10^{-21}$~erg~s$^{-1}$~cm$^{-2}$~arcsec$^{-2}$ that is fainter than our flux limit. The $f_\mathrm{ext}$ estimated in this study should thus be treated as the lower limit when comparing with the results from \citet{yuma17}, as our dataset may miss some diffused, spatially extended [O{\sc iii}] emission compared with \citet{yuma17}. With the shallower surface brightness limit, our results at $z=5-9$ are still $\sim1$~dex higher than those from \citet{yuma17}, indicating spatially extended [O{\sc iii}] emission is considerably more common in high-z galaxies. Such a redshift evolution of $f_\mathrm{ext}$, as shown in Figure \ref{fig:bfrac}, is consistent with the redshift evolution of halo merger rate that increases towards high redshift \citep{fakhouri10}. Contrastingly, the redshift evolution of $f_\mathrm{ext}$ deviates from the cosmic SFR density \citep[SFRD;][]{madau14}, which peaks at $z=2-3$ and decreases towards high $z$. Our results suggest that the spatial extension of [O{\sc iii}] is likely connected with outflow triggered by the more frequent halo mergers at high-$z$. The relatively large number fraction of spatially extended gas emission may also indicate that the post-outflow phase in the outflow cycle at high redshift has a longer duration that at low redshift, which is likely due to the low velocity of outflow wind in high-$z$ galaxies (Figure \ref{fig:vesc}).

Similarly, we estimate the broad line outflow fraction with our results in Section \ref{sec:outflow}, where we identify ongoing outflow in 3/18 and 2/12 galaxies at at $z=5.4-7.0$ and $7.0-8.9$, respectively. In the right panel of Figure \ref{fig:bfrac}, we compare our broad line outflow fraction with those at lower redshift traced by ionized gas. At $z\sim0$, \citet{wylezalek20} used SDSS-IV MANGA data to identified 257/3086 (8\%) galaxies have ongoing ionized gas outflow. At $z=0.6-2.7$, \citet{fs19} derived a outflow fraction of 25\% with 599 SF main sequence galaxies ranging from $10^{9.5}$ to $10^{11.5}~M_\odot$. Considering the errors due to the limited sample size, our results are comparable with the the $z\sim0$ and $z\sim2$ values, deviating from the evolution trend of major merger rate that increase rapidly towards high redshift. However, our results are derived from a parent sample with $M_*$ ranging from $10^7$ to $10^9~M_\odot$ that is smaller than the low-z studies. As outflow are more likely to occur in objects with higher stellar mass and/or SFR \citep[e.g.,][]{fs19}, our sample should be treated as the lower limit when comparing with the $f_\mathrm{out}$ values at lower redshift. We thus cannot rule out the possibility that $f_\mathrm{out}$ increases towards high redshift following the trend of major merger rate. 
Future observations on a larger sample of high-z galaxies with the high-resolution NIRSpec spectroscopy or IFU will improve the estimation in both $f_\mathrm{ext}$ and $f_\mathrm{out}$, which may reaveal the driving force of ongoing gas outflow and spatial extension.

\subsection{Origins of [O{\sc iii}] Spatial Extension and Ongoing Outflow}\label{subsec:blob_origin}
In Figure \ref{fig:sncut}, there are 13 objects locating on the top-right region that have both $S/N_\mathrm{img}>7$ and $S/N_\mathrm{spec}>30$. Among these 13 objects, there are one objects that only has spatially extended [O{\sc iii}] emission, three with only broad [O{\sc iii}] emission lines, and one with both features. This raises an interesting question on the discrepancy between the spatially extended [O{\sc iii}] gas and broad [O{\sc iii}] emission lines, as one would expect to observe both of these two features in galaxies with outflows. We discuss the possible scenarios that can cause such a discrepancy between the galaxies with spatially extension and broad emission lines of ionized gas. 

The first possibility is the different viewing angles on galaxies with biconical outflows, which has been observed in local galaxies \citep[e.g.,][]{venturi18} and recently proposed by \citet{carniani23} at high-$z$. Assuming a biconical outflow with an opening angle ($\theta$), the broad emission lines can only be observed in the directions of the outflow. The probability for observing the broad line components is approximated by the fraction of solid angle ($\Omega$) covered by the outflows, which is given by:
\begin{equation} \label{eq:open_angle}
    \Omega = 2\times [2\pi(1-\cos{\theta})],
\end{equation}
With $\Omega/4\pi = 4/13$ (see Figure \ref{fig:sncut}), we obtain $\theta = 46.2^\circ$ using Equation \ref{eq:open_angle}. Considering the errors associated with the small number counts of both galaxies and outflow objects, the range of $\theta$ is $29.8^\circ-58.9^\circ$, which is consistent with the results obtained in the local Universe \citep[e.g.,][]{venturi18}. However, it should be noticed that the observed outflow geometry is more complicated than the simple biconical model. In such a case, the objects with only spatially extended [O{\sc iii}] emission may be explained by the clumpy geometry of outflows in individual galaxies.

The second possibility is the gas cycle in the periodic SF history. We classify our galaxies into four evolution phases of outflow. In the first phase of early outflow, ionized gas is evacuated from the vicinity of SF regions due to stellar wind or AGN feedback but yet to reach the CGM of galaxies. This phase corresponds to the $3/14$ galaxies in our sample that show broad line outflows in NIRSpec spectra but without spatially extended gas emission in NIRCam images. As the outflowing gas extends to the CGM during the second phase of late outflow, we can now observe both broad line outflow as well as spatial extension of ionized gas, whose example is given by ERO\_05144. When gas outflow finally subsides, a spatially extended shell of ionized gas would form around the CGM of galaxies if the outflow wind is not energetic enough to escape from the DM halos of galaxies, which is the case for our high-$z$ galaxies shown in Section \ref{sec:outflow}. This marks the third phase, post-outflow, during which only spatially extended gas emission can be observed while broad line outflow is absent (i.e., $1/14$ galaxy in Figure \ref{fig:sncut} without broad line). Finally, the gas falls back to the SF region due to gravitation and fuels the SF activities that would initiate another cycle of gas outflow. This marks the last phase of inflow+starburst, which is like represented by the compact SF galaxies without spatially extended gas emission or broad lines. Such a cycle of gas outflow, also known as ``gas fountain'', is consistent with the predictions given by hydrodynamic simulations \citep[e.g.,][]{sd15,hh17}, as well as our results of weak outflows presented in Section \ref{sec:outflow}. While the typical SF and feedback time scale for high-$z$ galaxies predicted by previous hydrodynamic simulations is 10$^8$~yr \citep[e.g.,][]{yajima22}, it remains a challenge to constrain the time scale for the post-outflow and inflow/starburst phases due to various SF and feedback models. Albeit the small sample size, our estimation on the time scales of different phases of outflows may provide hints to future simulation works. We also notice that it is difficult to fully distinguish the degeneracy between the gas cycle and viewing angle scenarios \citep[see also][]{xu24}, which requires a larger number of galaxies in different phases and further constraints on their star-forming histories and/or dynamics.

Apart from these two scenarios, there are other possibilities that can cause the spatial extension and/or broad components of [O {\sc iii}] emission, such as merging events. We examined the morphology of our objects in the rest-frame UV images taken with the F115w or F150w filters with the highest spatial resolution. As shown in the left panals in Figure \ref{fig:z6b_img}, we do not find the features that indicate mergers, which suggests that the spatial extension and/or broadening of [O{\sc iii}] emission are unlikely due to mergers. However, the scenario of merging events cannot be fully excluded with the possibility that these objects may represent the very last stages of mergers. Future IFU observations that map the spatial distributions and dynamics of ionized gas would help distinguish these possible scenarios.

\subsection{Powering Sources}\label{subsec:blob_source}
\begin{figure*}[ht!]
\begin{center}
\includegraphics[scale=0.7]{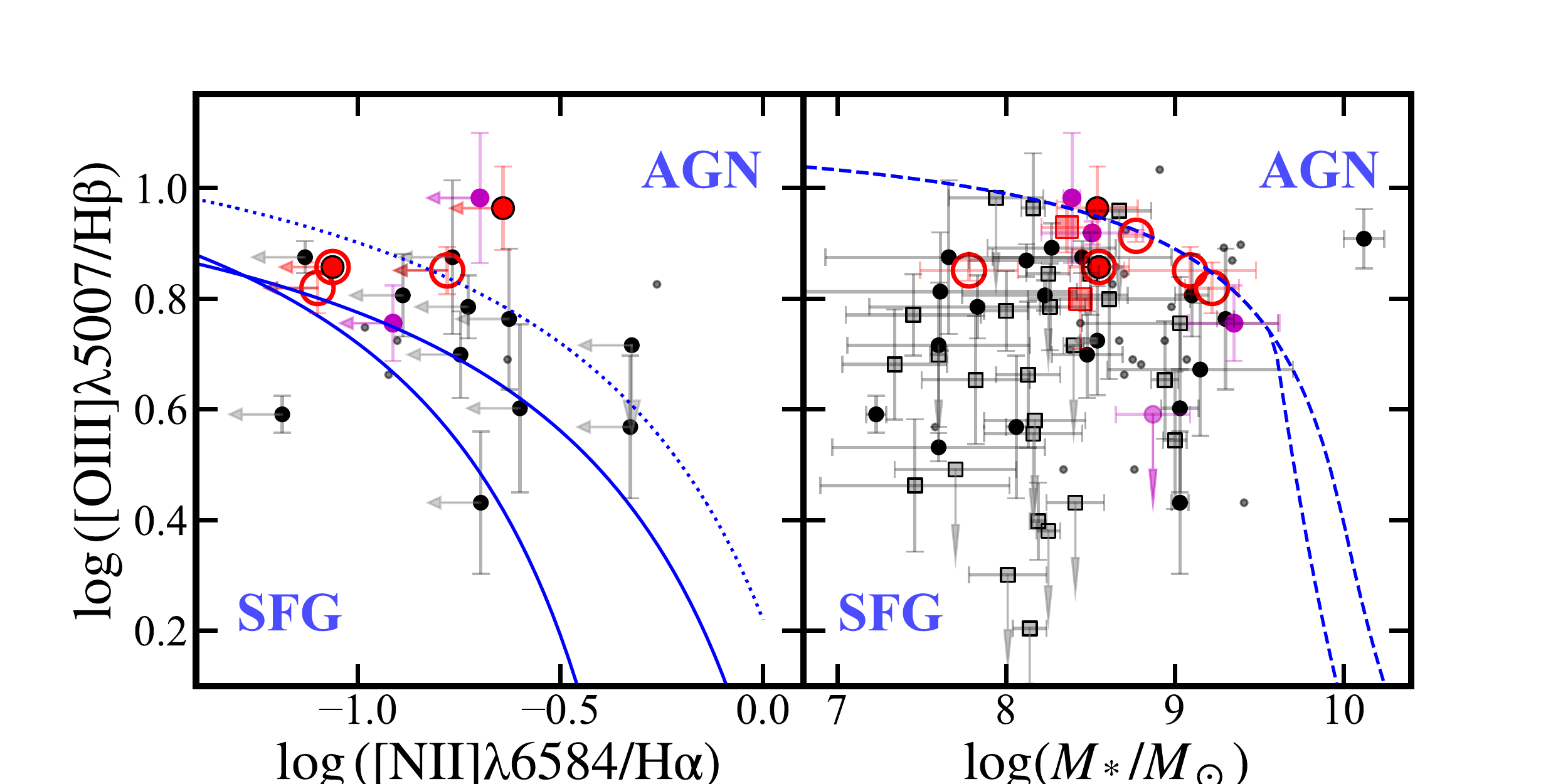}
\end{center}
\caption{Left: BPT diagram of our galaxies with [N{\sc ii}]+H$\alpha$ complex covered by NIRSpec. The data points are identical to Figure \ref{fig:ms_sfr}. The blue solid curves shows the SFG-AGN separation at the local universe determined by \citet{kewley01} and \citet{kauffmann03}. The blue dotted curve is the SFG-AGN separation at $z\sim3$ obtained by \citet{kewley13}. Right: MEx diagram of our galaxies with data points identical to Figure \ref{fig:ms_sfr}. The blue dashed curves indicate the SFG-AGN separation obtained by \citet{juneau14}. At high-$z$, this separation curve is expected to shift towards lower $M_*$ due to the redshift evolution of $M_*$.}\label{fig:bpt}
\end{figure*}

There are several plausible origins that can trigger the gas outflow traced by spatial extension and/or broad lines of [O{\sc iii}] emission. At low redshift, both SF galaxies and AGN have been found to drive the [O{\sc iii}] spatial extension \citep[e.g.,][]{yuma17} and ongoing outflow \citep[e.g.,][]{zakamska16,vayner21}.
We first examine the existence of type~1 AGN in the nine objects that have ongoing [O{\sc iii}] outflow and/or spatial extension (Section \ref{subsec:sample} and \ref{sec:outflow}) by checking their H$\beta$ and H$\alpha$ emission lines. We find that none of these nine objects have broad components (FWHM$\geq 1000$~km~s$^{-1}$) in H$\beta$ and H$\alpha$ emission lines that are similar to previously reported type~1 AGN identified in JWST NIRSpec spectra \citep[e.g.,][]{harikane23b,kocevski23,larson23,ubler23}. The absent of broad components in Balmer lines suggest that our ongoing [O{\sc iii}] outflow and/or spatial extension are not caused by type~1 AGN. 
Next, we investigate the existence of type~2 (obscured) AGN with two typical diagnostics, BPT and mass-excitation (MEx) diagram. In the left panel of Figure \ref{fig:bpt}, we plot the six galaxies that have [O{\sc iii}] spatial extension and/or ongoing outflow and with [N{\sc ii}]+H$\alpha$ complex covered by the wavelength range of NIRSpec. Although these six galaxies have relatively high [O{\sc iii}]/H$\beta$ line ratios of $6.3-10.0$, we can only obtain the upper limits of [N{\sc ii}]/H$\alpha$ line ratio, indicating that they can be either SF galaxies or AGN. For the MEx diagram (right panel of Figure \ref{fig:bpt}), most of the nine galaxies with [O{\sc iii}] spatial extension and/or ongoing outflow locate in the SF galaxy regime but close to the curve that separates the SF galaxy and AGN regimes. It should also be noted that the MEx diagram was calibrated with galaxies at $z<2$, and has been found to mis-classify AGN with low masses and SF galaxies with high ionization parameters at higher redshift \citep[e.g.,][]{coil15}. Furthurmore, with the redshift evolution of $M_*$ the separation curve of SF galaxy and AGN is expected to shift towards the low-$M_*$ side. With these potential caveats and biases, we cannot rule out the existence of obscured (type~2) AGN in these galaxies. It is also possible that the spatially extended [O{\sc iii}] and ongoing outflow are the remnants of past AGN activity. In such cases, previous AGN episodes initiated the ionized gas outflow and subsided while the outflow signatures are still observable \citep[e.g.,][]{ishibashi15,wylezalek20}. 

\section{Summary} \label{sec:summary}
We investigate the spatial extension and broad line outflow of ionized gas traced by rest-frame optical [O{\sc iii}] emission in 61 galaxies at $z=5.4-8.9$ that are confirmed by NIRSpec data and covered by NIRCam footprints. Locating around the $M_*-$SFR main sequence, these galaxies represent typical star forming galaxies with $M_*=10^{7-10}M_\odot$. Utilizing NIRCam multiband imaging data, we construct [O{\sc iii}] emission line images by subtracting the stellar continua from the broadband images that contain [O{\sc iii}] emission. Our results are summarized below:

We identify 4/11 and 0/7 galaxies at $z=5.4-7.0$ and $7.0-8.9$, respectively, have [O{\sc iii}] emission spatially extended beyond their stellar continua at kpc scale. We do not find significant trends in the $M_*-$SFR distribution of these galaxies, indicating the complex physical origins of [O{\sc iii}] spatial extension. Two of these four galaxies were observed with the low-resolution Prism with $R\sim100$, which is too low to determine the existence of ongoing outflow through spectral fitting. Among the other two galaxies observed with medium-resolution filter-grating pairs, only one galaxy, ERO\_05144, has ongoing gas outflow features suggested by a second broad Gaussian component of [O{\sc iii}] emission lines in NIRSpec. 

We search for broad line [O{\sc iii}] outflow by fitting the H$\beta$ and [O{\sc iii}]$\lambda\lambda$4059,5007 emission lines of 30/61 galaxies that were observed with medium- or high-resolution filter-grating pairs, obtaining a total of five galaxies with ongoing [O{\sc iii}] outflows indicated by a second broad Gaussian components in [O{\sc iii}]$\lambda\lambda$4059,5007 emission lines. Apart from ERO\_05144 that has spatial extension, the other four galaxies do not show spatially extended [O{\sc iii}] emission. All of the five galaxies with ongoing outflow have $v_\mathrm{max}/v_\mathrm{esc}$ values smaller than unity, indicating that the ionized gas ouflows are unlikely to escape from the DM halos. 

We derive the number fraction of spatially extended [O{\sc iii}] emission and broad line outflow. The number fractions of spatially extended [O{\sc iii}] emission at $z\sim5-9$ are considerably higher than the local values due to the higher merger rate in the earlier epochs, suggesting that the outflow activities in high-$z$ galaxies may be triggered by the frequent major mergers events, or that the post-outflow phase in the gas cycle has a longer duration for high-$z$ galaxies.

The discrepancy between galaxies with spatially extended [O{\sc iii}] emission and with broad line outflow can be explained by either the different viewing angles towards complicated outflow geometries, or the gas cycle in a periodic SF history of early, late, and post-outflow phases, during which the ionized gas is first expelled from the vicinity of SF regions, extending to the CGM while losing their energy, and being trapped in the CGM after outflow phases subsides due to the low outflow velocity. Such a gas cycle is likely driven by stellar wind, although the possibilities of type~2 AGN and previous AGN activities cannot be excluded. 

\section{Acknowledgments}
We thank the anonymous referee for constructive comments and suggestions, and H. Yajima for the support towards the completion of this work. This work is based on observations made with the NASA/ESA/CSA James Webb Space Telescope. The data were obtained from the Mikulski Archive for Space Telescopes at the Space Telescope Science Institute, which is operated by the Association of Universities for Research in Astronomy, Inc., under NASA contract NAS 5-03127 for JWST. These observations are associated with programs ERS-1324 (GLASS), ERS-1345 (CEERS), GO-2561 (UNCOVER), and ERO-2736. 
The authors acknowledge the teams of JWST commissioning, ERO, GLASS, UNCOVER, and CEERS led by Klaus M. Pontoppidan, Tommaso Treu, Ivo Labbe/Rachel Bezanson, and Steven L. Finkelstein, respectively, for developing their observing programs with a zero-exclusive-access period. The observations analyzed in this work can be accessed via \dataset[10.17909/qp74-bk09]{https://doi.org/10.17909/qp74-bk09}. This work is supported by the World Premier International Research Center Initiative (WPI Initiative), MEXT, Japan, and KAKENHI (JP23KJ0589, 20H00180, 21H04467, and 21J20785) through Japan Society for the Promotion of Science. Y.Z. acknouledges the support from JST SPRING (JPMJSP2108), as well as the joint research program of the Institute for Cosmic Ray Research (ICRR), University of Tokyo.

%



\software{astropy \citep{2013A&A...558A..33A,2018AJ....156..123A},  
          Source Extractor \citep{1996A&AS..117..393B}
          }




\bibliography{sample631}{}
\bibliographystyle{aasjournal}



\end{document}